\pdfoutput=1
\documentclass{PoS}

\usepackage{amsmath}
\usepackage{slashed}
\usepackage{amssymb}
\usepackage{euscript}
\usepackage{bbm}
\usepackage{bm}
\usepackage[T1]{fontenc}
\usepackage[utf8]{inputenc}
\usepackage{graphicx}

\newcommand{\bk}{{\bf{k}}}

\title{Monte Carlo modelling of NLO DGLAP QCD Evolution in the fully unintegrated form}

\ShortTitle{
Monte Carlo modelling of NLO DGLAP QCD Evolution in the fully unintegrated form}

\author{\speaker{A. Kusina}%
        \thanks{This work is partly supported by the EU 
  			Framework Programme grants MRTN-CT-2006-035505 
			and by the Polish Ministry of Science and Higher Education grants
  			No. 1289/B/H03/2009/37 and 153/6.PR UE/2007/7.}\\
         H. Niewodniczański Institute of Nuclear Physics, Polish
  	Academy of Sciences, \\ ul. Radzikowskiego 152, 31-342 Krakow, Poland\\
       E-mail: \email{Aleksander.Kusina@ifj.edu.pl}}

\author{S. Jadach\\
         H. Niewodniczański Institute of Nuclear Physics, Polish
  	Academy of Sciences, \\ ul. Radzikowskiego 152, 31-342 Krakow, Poland\\
        E-mail: \email{Stanislaw.Jadach@ifj.edu.pl}}

\author{M. Skrzypek\\
         H. Niewodniczański Institute of Nuclear Physics, Polish
  	Academy of Sciences, \\ ul. Radzikowskiego 152, 31-342 Krakow, Poland\\
        E-mail: \email{Maciej.Skrzypek@ifj.edu.pl}}

\author{M. Slawinska\\
        H. Niewodniczański Institute of Nuclear Physics, Polish
  	Academy of Sciences,\\ ul. Radzikowskiego 152, 31-342 Krakow, Poland\\
        E-mail: \email{Magdalena.Slawinska@ifj.edu.pl}}

\abstract{%
Presently available perturbative QCD calculations combining
hard process matrix element with the Parton Shower Monte Carlo programs
feature hard process matrix element calculated often
beyond the leading order (LO),
that is including complete next-to-leading-order (NLO),
or even next-to-next-to-leading-order (NNLO) corrections,
while Parton Shower is only at the leading order (LO).
We report here on a work in progress which demonstrate feasibility
of constructing Parton Shower Monte Carlo (PSMC) featuring complete
NLO corrections to QCD evolution with respect to the logarithm of the
factorization scale.
This effort presently covers non-singlet subset of Feynman diagrams contributing
to the above QCD evolution.
It should be stressed that our approach to the NLO QCD evolution is {\em exclusive},
that is giving insight into the fully unintegrated phase space.
However, at the inclusive level our implementation
agrees exactly with the standard inclusive picture of the NLO DGLAP evolution.
Our new approach
(after including NLO singlet diagrams)
provides a complete method of combining the resummed and fixed order
perturbative QCD calculations beyond the LO in a form suitable for the MC
implementation, an alternative to the existing ones.
First practical applications will include
Monte Carlo generators for W/Z production processes in hadron-hadron colliders
and lepton-hadron colliders.}

\FullConference{35th International Conference of High Energy Physics\\
          July 22-28, 2010\\
          Paris, France}

\begin{document}

As the Large Hadron Collider (LHC) just started to operate,
experiments will need high accuracy QCD calculations
not only for inclusive observables,
but also for multiparton final states,
especially in a form of events from Parton Shower Monte Carlo (PSMC) programs.
Huge range of energy scales in the PSMCs will require
including beyond the LO calculations not only in the hard process part
but also in the middle of the shower (ladder).
Why 25 years old traditional PSMCs,
see refs.~\cite{PSMCs},
are still restricted to the LO level only?
Main reasons are that:
(A) principal role of PSMCs was to hadronize quark and gluons into hadrons,
(B) hadron collider data were until recently rather poor,
(C) conceptual problem on the side of the factorization theorems
of QCD~\cite{FACth}, badly suited for the MC implementation, as they
(i) violate 4-momentum conservation,
(ii) irreversibly integrate over real emissions
(iii) feature huge over-subtractions and cancellations
(iv) operate with non-positive distributions,
(D) last not least computers were until recently to slow to consider
unintegrated NLO corrections inside the PSMC ladder.
With the presented works an era of NLO level PSMCs finally comes soon!
Possible profits/gains from NLO PSMC are the following:
(a) new simpler and cleaner schemes of
matching of the hard process ME at NLO and NNLO level with NLO PSMC,
(b) natural extensions towards BFKL/CCFM evolution at low $x$,
(c) better modelling of low scale phenomena, $Q<10GeV$,
(d) new MC tools for porting information on parton distributions from
one process to another (from DIS at HERA to W/Z production at LHC)
and from one scale to another -- an alternative for traditional inclusive PDFs.
In our work presented here,
we take scheme of Curci, Furmanski, Petronzio (CFP)~\cite{Curci:1980uw}
as a guide and we concentrate on the non-singlet NLO evolution
for the initial state radiation (ISR) of multiple gluons.
NLO corrections are added to each LO kernel in the middle of LO
ladder of the PSMC, see below,
\begin{center}
  {\includegraphics[height=40mm]{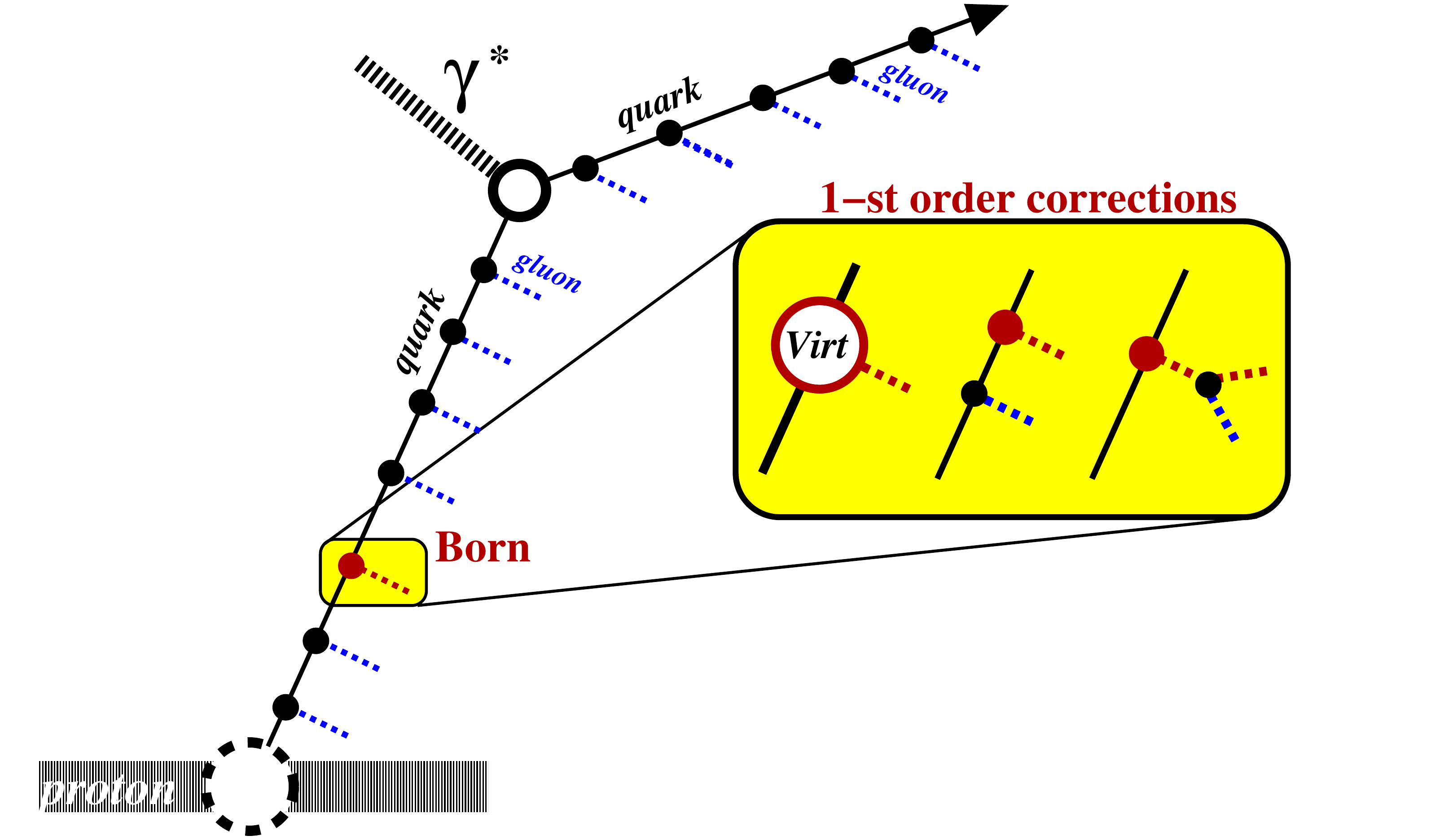}}
\end{center}
such that at the inclusive level we reproduce NLO DGLAP~\cite{DGLAP} evolution.
As depicted above, our starting point is the LO ISR single ladder like in DIS.
Let us start with including NLO corrections
to the gluon emission vertex at the top of this LO ladder,
\begin{equation}
\begin{split}
&\bar{D}^{[1]}_{B}(x,Q)=
e^{-S_{_{ISR}}}
\raisebox{-35pt}{\includegraphics[height=30mm]{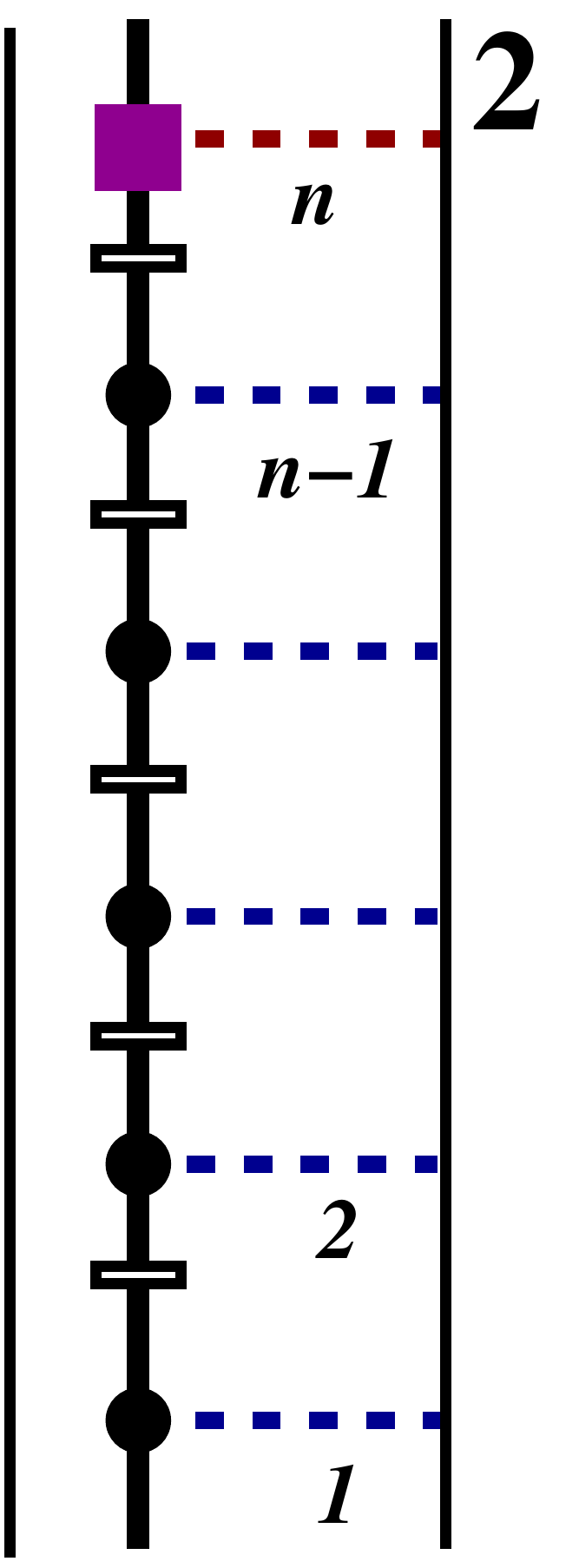}}
\;\; + e^{-S_{_{ISR}}}
\sum\limits_{j=1}^{n-1}
\raisebox{-35pt}{\includegraphics[height=30mm]{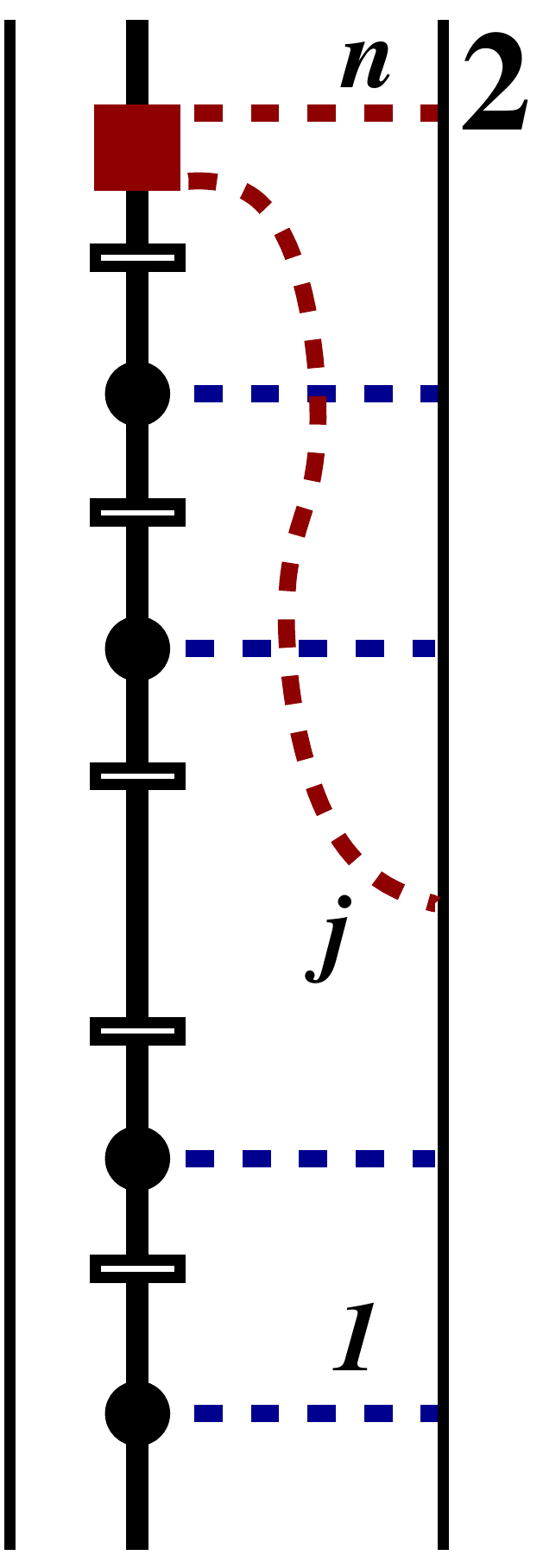}}
= e^{-S_{_{ISR}}}
\Bigg\{ \delta_{x=1}+
\\&~~~~~~~
+\sum_{n=1}^\infty\;
\bigg( \prod_{i=1}^n\; 
    \int\limits_{Q>a_i>a_{i-1}}\!\!\!\!\!\!  \frac{d^3 k_i}{k_i^0}\;
    \rho^{(1)}_{1B}(k_i)
\bigg)
\bigg[ \beta_0^{(1)}(z_n)
+ \sum_{j=1}^{n-1}W(\tilde{k}_n, \tilde{k}_j)
\bigg]
\delta_{x=\prod_{j=1}^n x_j}
\Bigg\},
\end{split}
\label{eq:NLO1}
\end{equation}
where the Monte Carlo weights reweighting LO distributions to NLO level is given by:
\begin{equation}
\beta_0^{(1)}=\frac{%
 \left| \raisebox{-7pt}{\includegraphics[height=7mm]{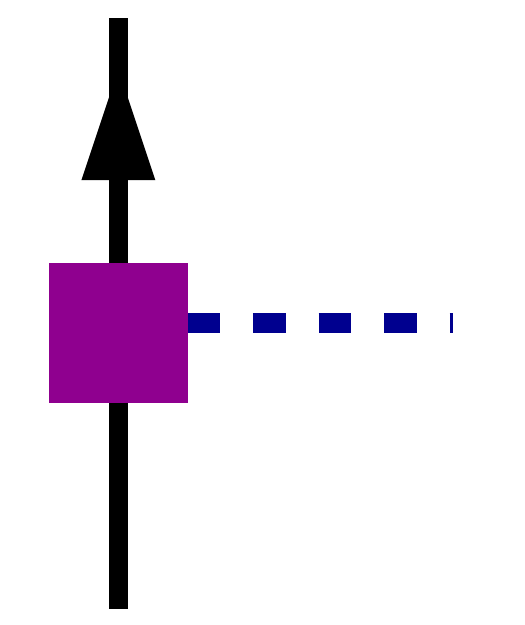}}
 \right|^2}%
{ \left|\raisebox{-7pt}{\includegraphics[height=7mm]{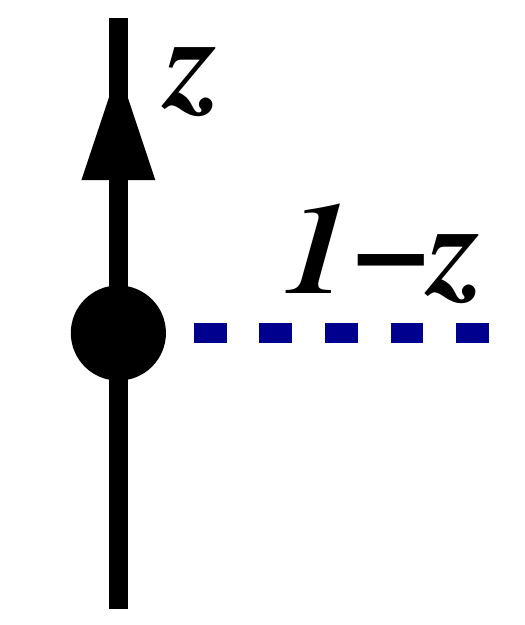}}
 \right|^2
},\quad
W(k_2,k_1)=\frac{%
 \left| \raisebox{-7pt}{\includegraphics[height=7mm]{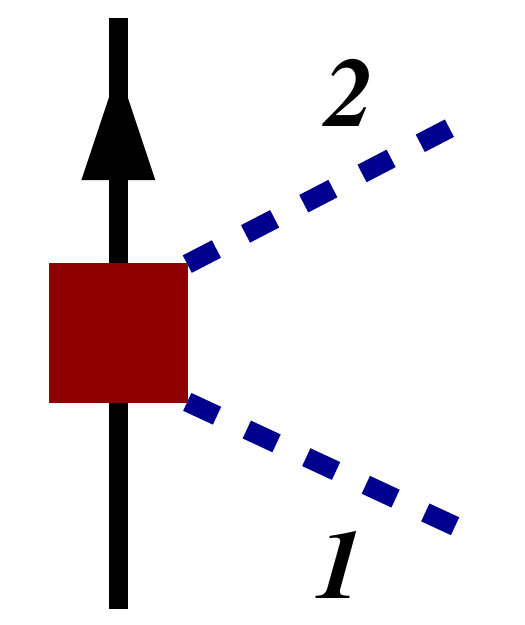}}
 \right|^2
}{%
 \left| \raisebox{-7pt}{\includegraphics[height=7mm]{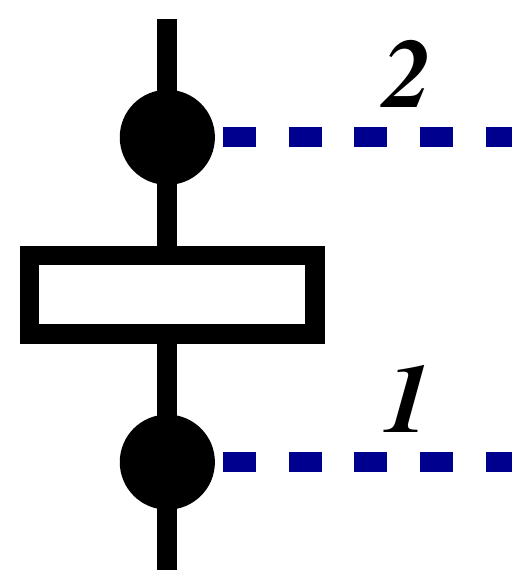}}
 \right|^2
}=
\frac{%
 \left| \raisebox{-7pt}{\includegraphics[height=7mm]{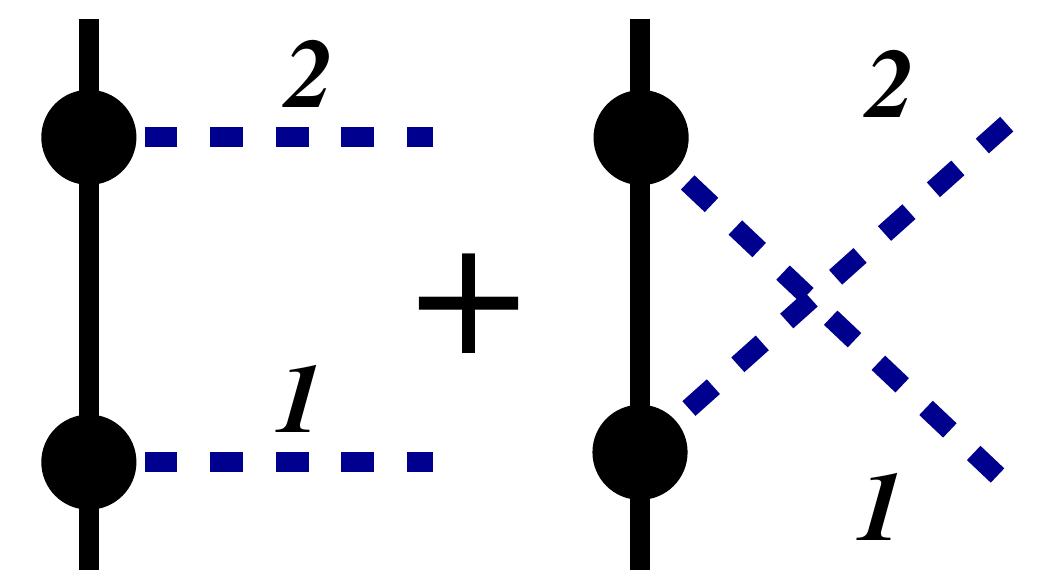}}
\right|^2
}{%
 \left| \raisebox{-7pt}{\includegraphics[height=7mm]{xBr2ReCt.pdf}}
 \right|^2
}\; -1.
\end{equation}
Let us explain first all parts and the ingredients in Eq.~\eqref{eq:NLO1}.
First term (ladder) is just the LO ladder
multiplied by NLO virtual correction factor for the last vertex,
where $\rho^{(0)}_{1B}(k_i)=\frac{2C_F^2\alpha_s}{\pi}\frac{1}{k_i^{T2}}
\frac{1+z^2}{2}$ is LO Altarelli-Parisi evolution kernel,
$\theta_{Q>a_i>a_{i-1}}$ indicates the ordering in angle ($a_i$ is the
rapidity related variable defined as $a_i=|\bk_{i\perp}|/\alpha_i$,
$\alpha_i$ is plus variable in Sudakov parametrization),
$\frac{ d^3 k_i}{k_i^0}$ is the Lorentz invariant phase-space and
$e^{-S_{_{ISR}}}$ is the Sudakov formfactor.
The NLO virtual correction
$\left|
 \raisebox{-7pt}{\includegraphics[height=7mm]{xBrBet0ISR.pdf}}
\right|^2\!\!\!
=\big(1+2\Re(\Delta_{_{ISR}}^{(1)})\big)\!
\left|
  \raisebox{-7pt}{\includegraphics[height=7mm]{xBrBorn.pdf}}
\right|^2$
simply multiply the LO Born (last) vertex and since it is finite,
it can be easily implemented by reweighting the LO distribution.
The $C_F^2$ real corrections coming from the subtracted bremsstrahlung type
diagrams
$
\left|
  \raisebox{-8pt}{\includegraphics[height=8mm]{xBrBetISR.pdf}}
\right|^2
=\left|
  \raisebox{-8pt}{\includegraphics[height=8mm]{xBrem2Real.pdf}}
\right|^2
-\left|
  \raisebox{-8pt}{\includegraphics[height=8mm]{xBr2ReCt.pdf}}
\right|^2
$
with the soft counterterm
$\left| \raisebox{-7pt}{\includegraphics[height=7mm]{xBr2ReCt.pdf}}
\right|^2$
subtracting the LO component affect pairs of partons (act in two
particle phase-space)%
\footnote{In our graphical notation the amplitude square means
cut diagrams for both Feynman graph and the counterterm.}.
The distinctive feature of Eq.~\eqref{eq:NLO1} is the presence
of the summation over trailing gluons along the ladder.
Without this summation (taking only $W(k_n,k_{n-1})$)
we wouldn't have the full NLO correction.

Including the full
NLO correction all over the ladder requires to sum over the positions 
of the real NLO insertions (corrections) in the ladder
and adding virtual corrections to all vertices:
\begin{equation}
\label{eq:D1B}
\begin{split}
&\bar{D}^{[1]}_{B}(x,Q)=
e^{-S_{_{ISR}}}
\sum\limits_{n=0}^{\infty}
\Bigg\{
\raisebox{-30pt}{\includegraphics[height=30mm]{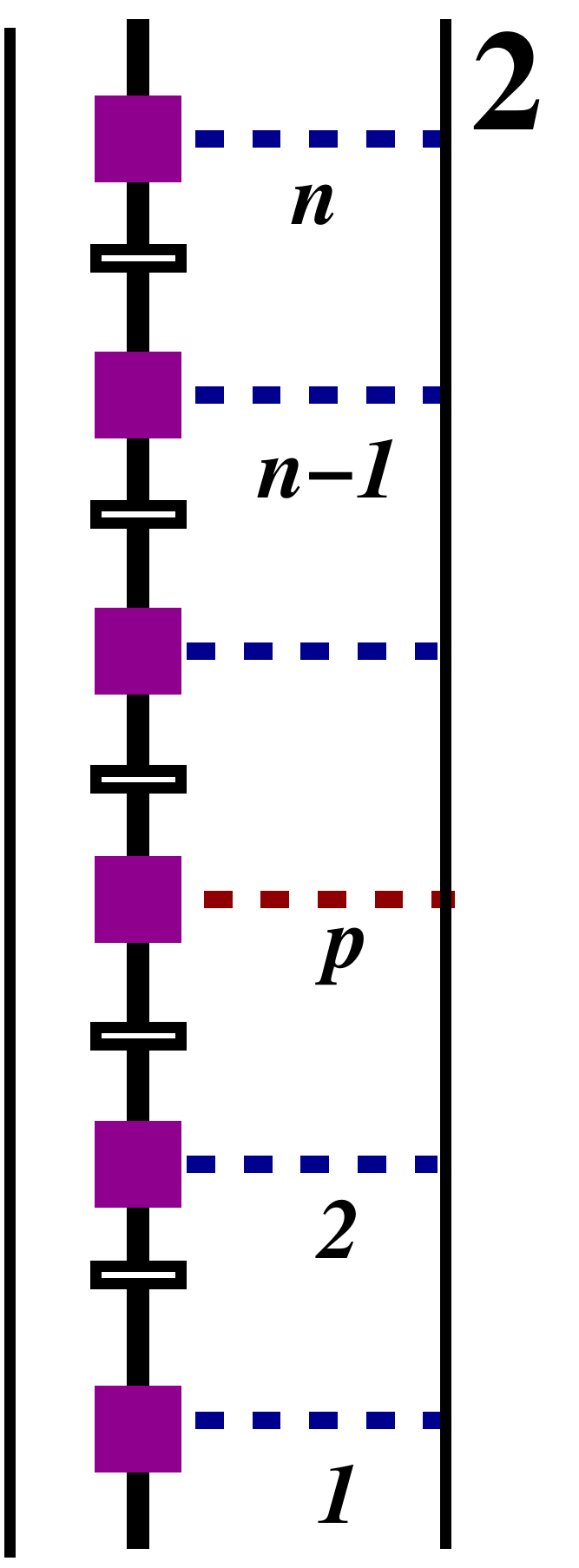}}
\;\; +
\sum\limits_{p_1=1}^{n}
\sum\limits_{j_1=1}^{p_1-1}
\raisebox{-30pt}{\includegraphics[height=30mm]{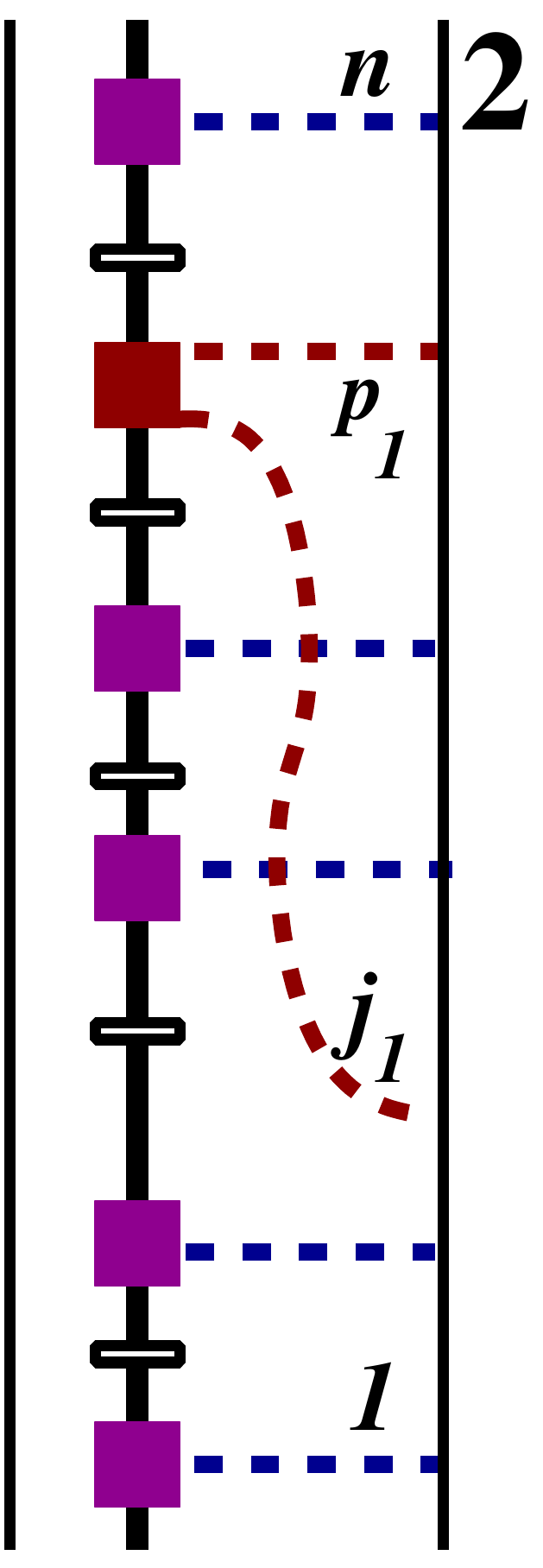}}
\;\; +
\sum\limits_{p_1=1}^{n}
\sum\limits_{p_2=1}^{p_1-1}
\sum\limits_{j_1=1 \atop j_1\neq p_2}^{p_1-1}
\sum\limits_{j_2=1 \atop j_2\neq p_1,j_2}^{p_2-1}
\raisebox{-30pt}{\includegraphics[height=30mm]{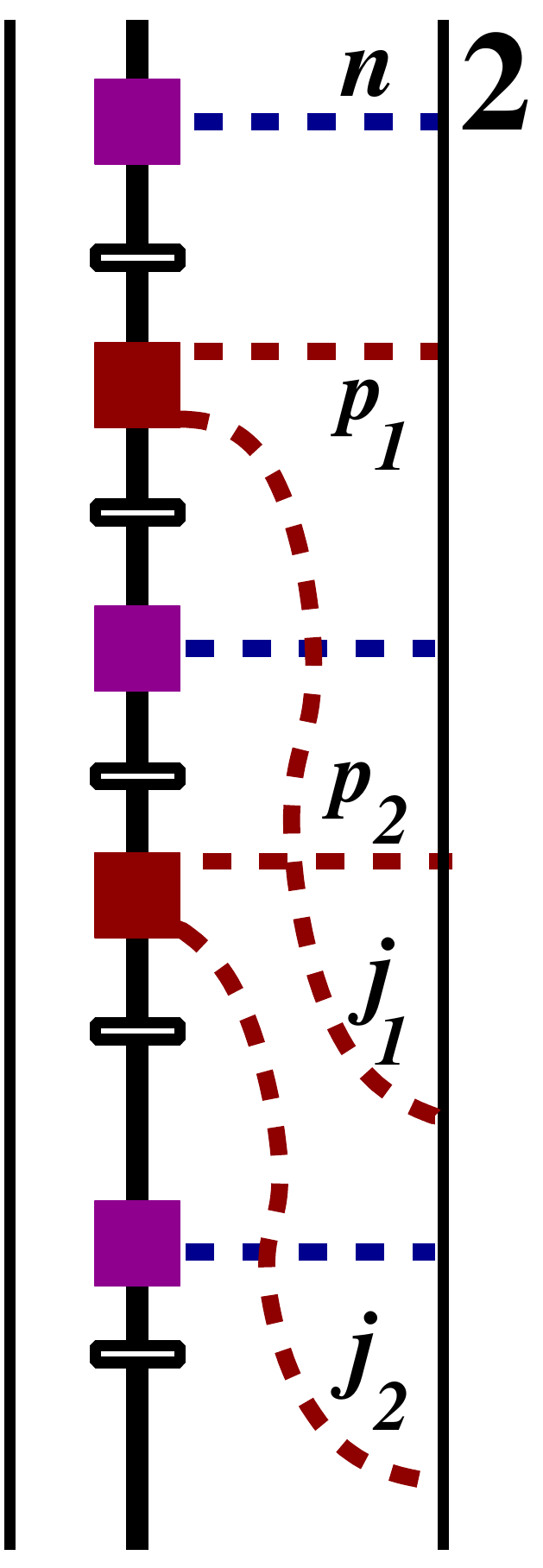}}+\dots
\Bigg\}
\\&
= e^{-S_{_{ISR}}}
\Bigg\{ \delta_{x=1}
+\sum_{n=1}^\infty\;
\bigg( \prod_{i=1}^n\; 
    \int\limits_{Q>a_i>a_{i-1}}\!\!\!\!\!\!  \frac{d^3 k_i}{k_i^0}\;
    \rho^{(1)}_{1B}(k_i) \beta_0^{(1)}(z_p)
\bigg)
\bigg[
1 + \sum\limits_{p=1}^{n} \sum_{j=1}^{p-1}
    W(\tilde{k}_p, \tilde{k}_j)+
\\&~~~~~~~~~~~~~
+\sum\limits_{p_1=1}^{n}
\sum\limits_{p_2=1}^{p_1-1}
\sum\limits_{j_1=1 \atop j_1\neq p_2}^{p_1-1}
\sum\limits_{j_2=1 \atop j_2\neq p_1,j_2}^{p_2-1}
W(\tilde{k}_{p_1}, \tilde{k}_{j_1})
W(\tilde{k}_{p_2}, \tilde{k}_{j_2})
+\dots
\bigg]
\delta_{x=\prod_{j=1}^n x_j}
\Bigg\},
\end{split}
\end{equation}
The above scheme is tested numerically
with 3-digit precision using prototype MC program:
\begin{center}
\includegraphics[width=70mm,height=45mm]{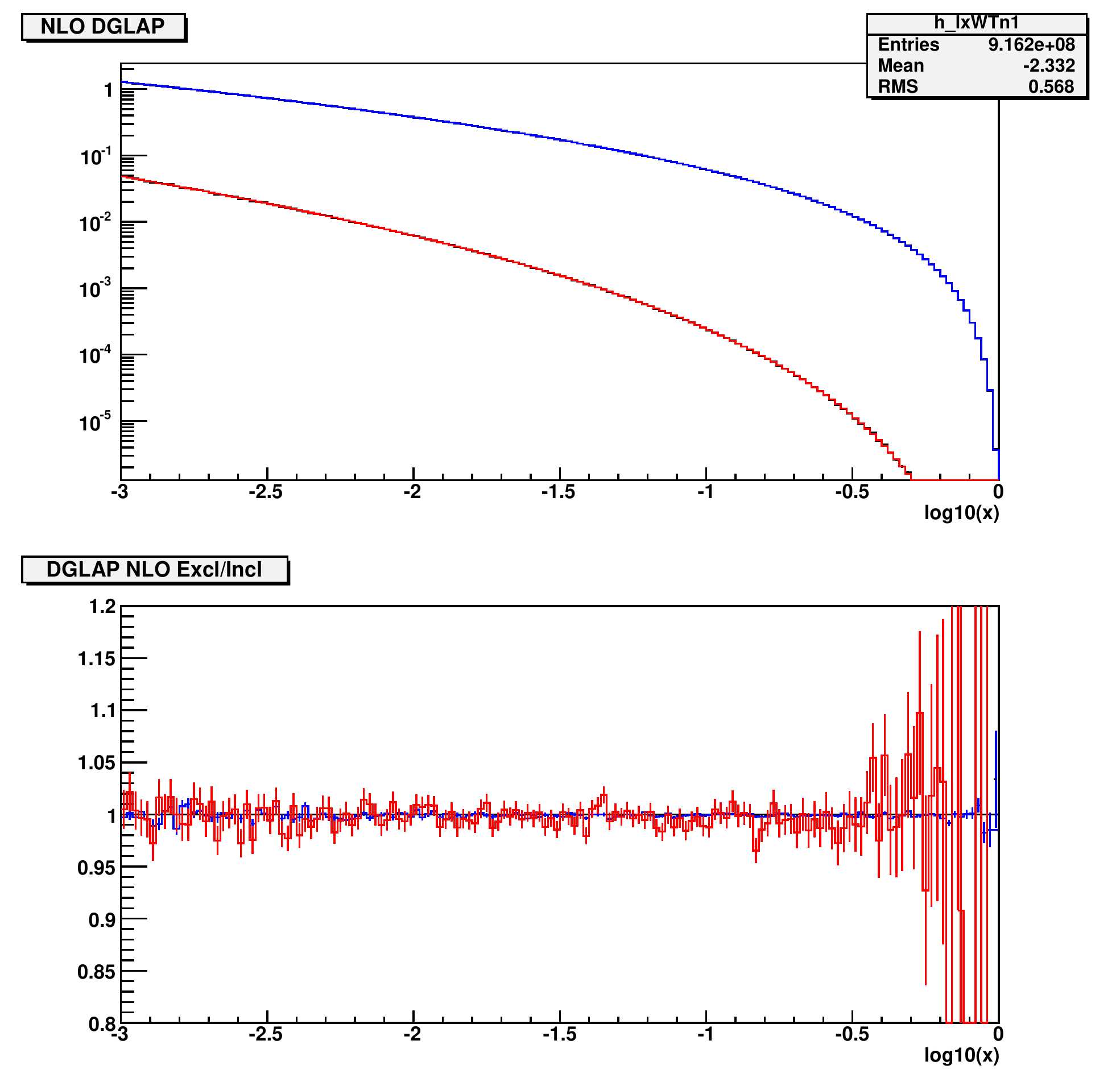}
\includegraphics[width=80mm,height=40mm]{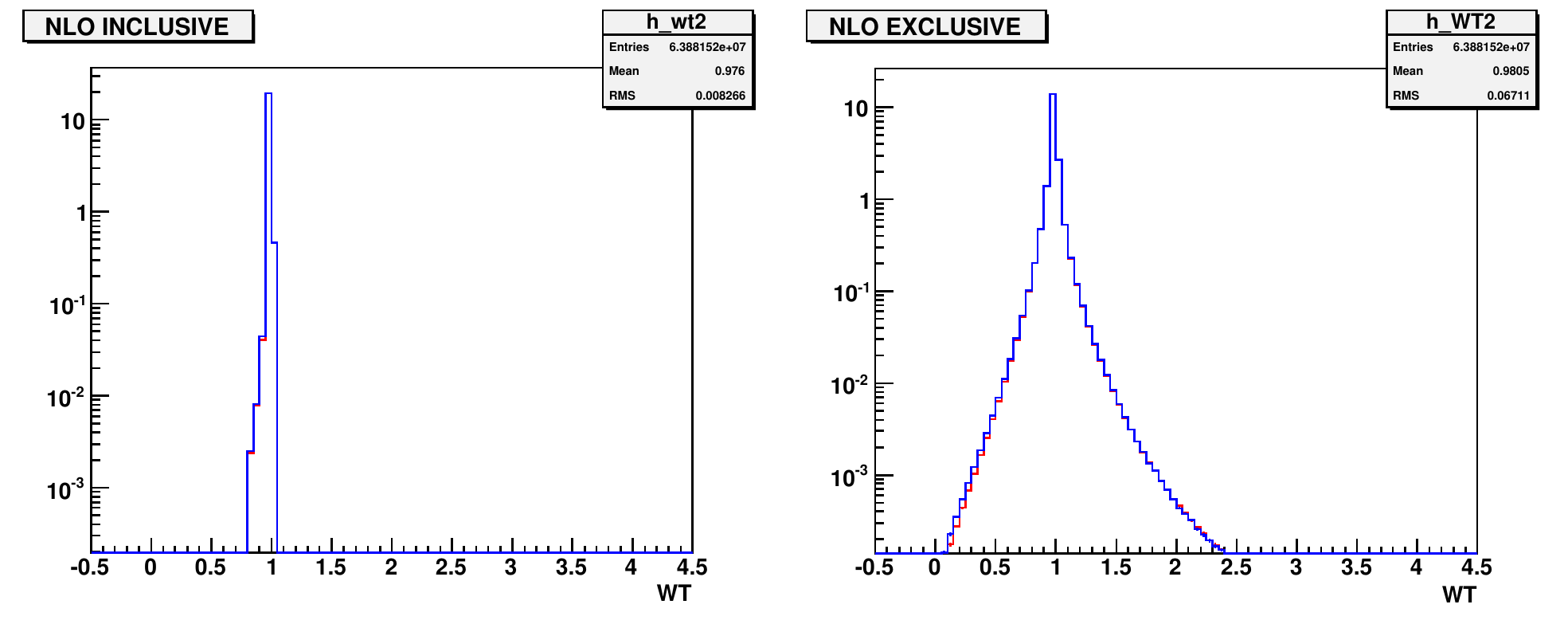}
\end{center}
Perfect agreement of the new scheme and traditional inclusive
approach is seen, especially in the left lower plot, where the ratios
for two sums in Eq.~(\ref{eq:D1B}) are plotted separately.
In the right plot we see the weight distributions for inclusive
and exclusive approach. The exclusive weight is positive and narrow.

In NLO corrections proportional to $C_AC_F$ color factor,
we get pair production diagram
$\left|
  \raisebox{-7pt}{\includegraphics[height=7mm]{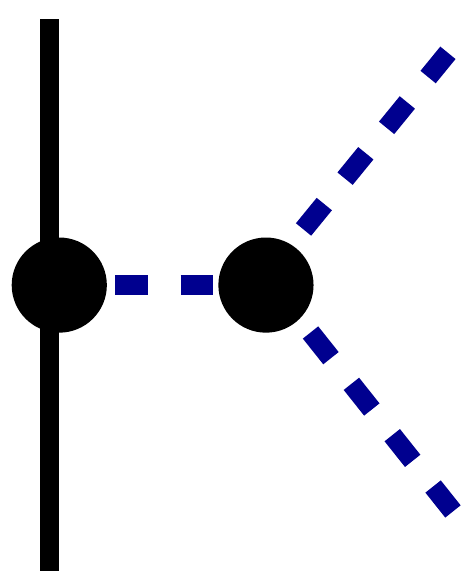}}
\right|^2$.
First question is why we treat this type of corrections in a special
way not just include them in the same manner as before.
The problem is that it induces Sudakov
double logarithm $+S_{FSR}$ (additional singularity when
produced pair is collinear) which would ruin the MC weight,
if we would restrict ourself to strict NLO only.
Resummation/exponentiation for this contribution is mandatory,
and new LO ladder on the external leg has to be implemented in the MC --
each emitted gluon will cascade into gluon emission ladder.
The FSR counterterm
$\left|
\raisebox{-7pt}{\includegraphics[height=7mm]{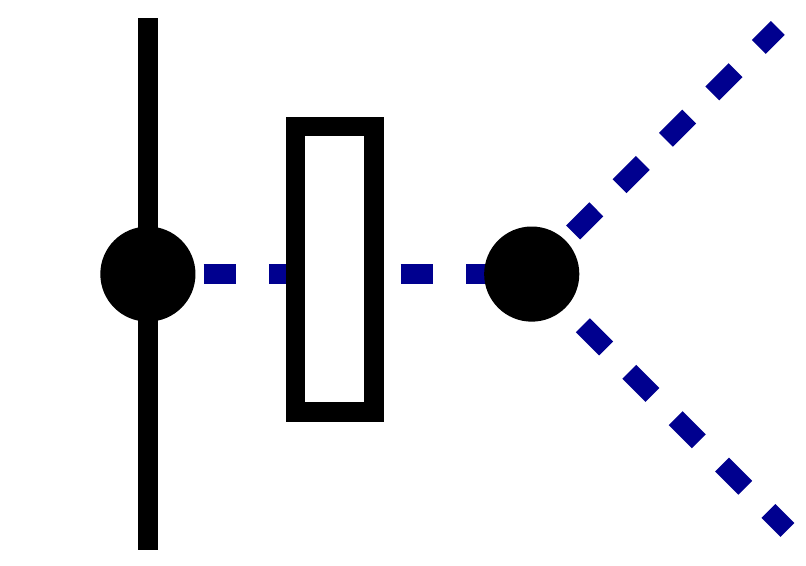}}
\right|^2$
is constructed, encapsulating
exactly soft and collinear singularities of $C_AC_F$ diagrams (both double and single logs).
The expression for the MC distribution with the last emission
in the ladder upgraded from the LO to NLO level reads:
\begin{equation}
\begin{split}
&~\!\!\!\!\!\!\!\!\!\!\!\!
\bar{D}^{[1]}_{NS}(x,Q)=
\\&
~\!\!\!\!\!\!\!\!\!\!\!\!
e^{-S}
\sum\limits_{n,m=0}^{\infty}
\Bigg\{
\raisebox{-25pt}{\includegraphics[height=25mm]{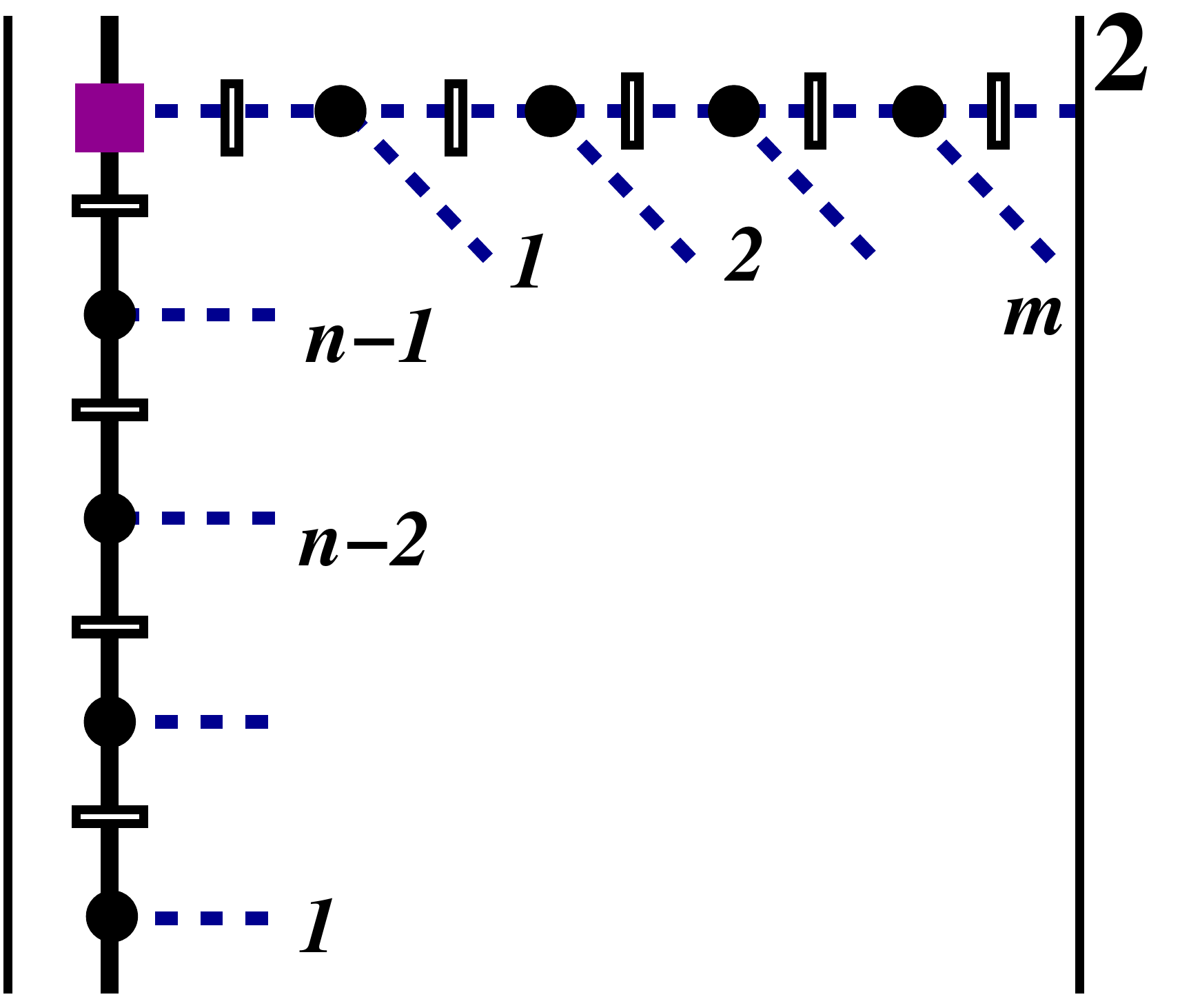}}
+\sum\limits_{j=1}^{n-1}
\raisebox{-25pt}{\includegraphics[height=25mm]{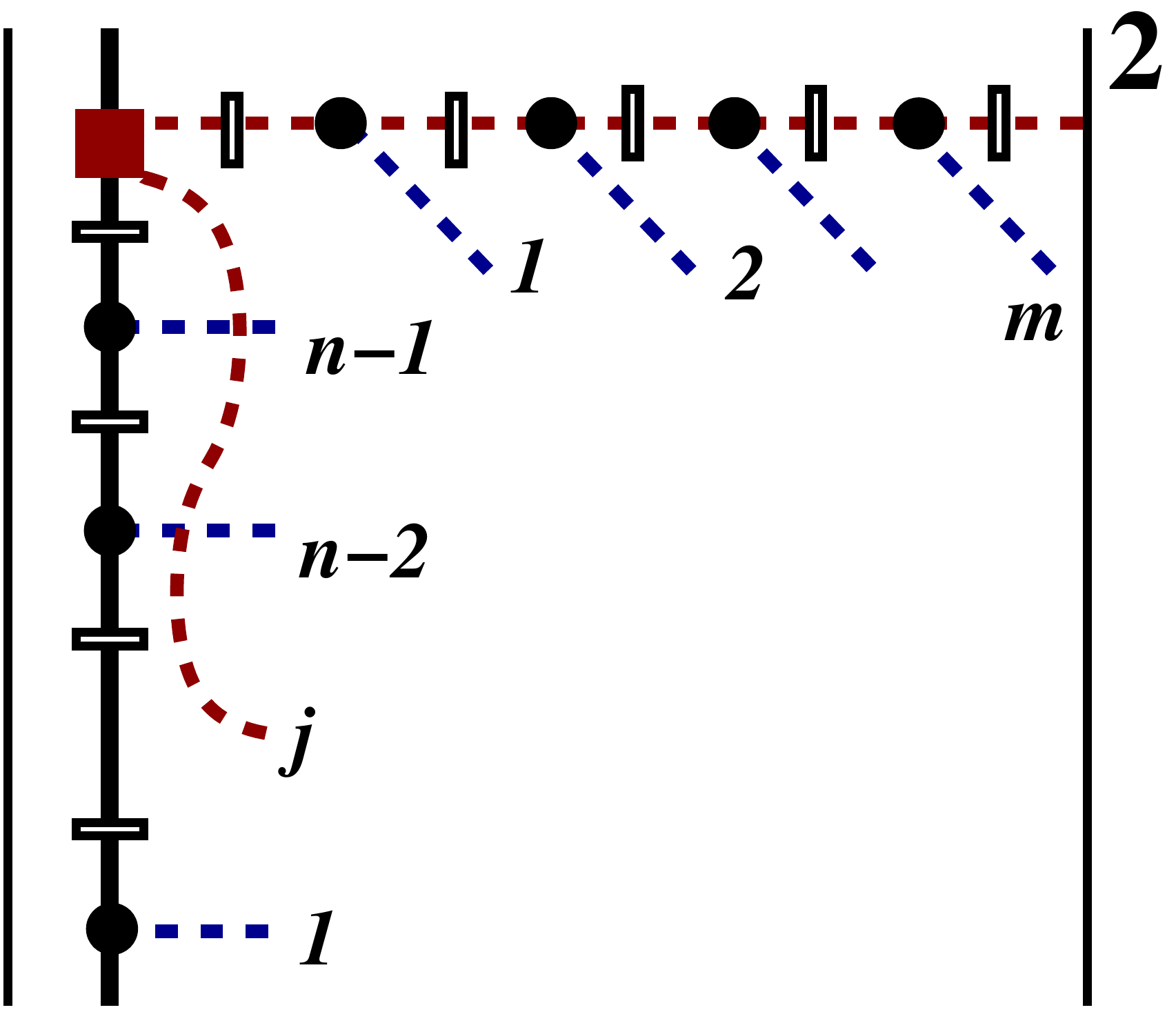}}
+\sum\limits_{r=1}^{m}
\raisebox{-25pt}{\includegraphics[height=25mm]{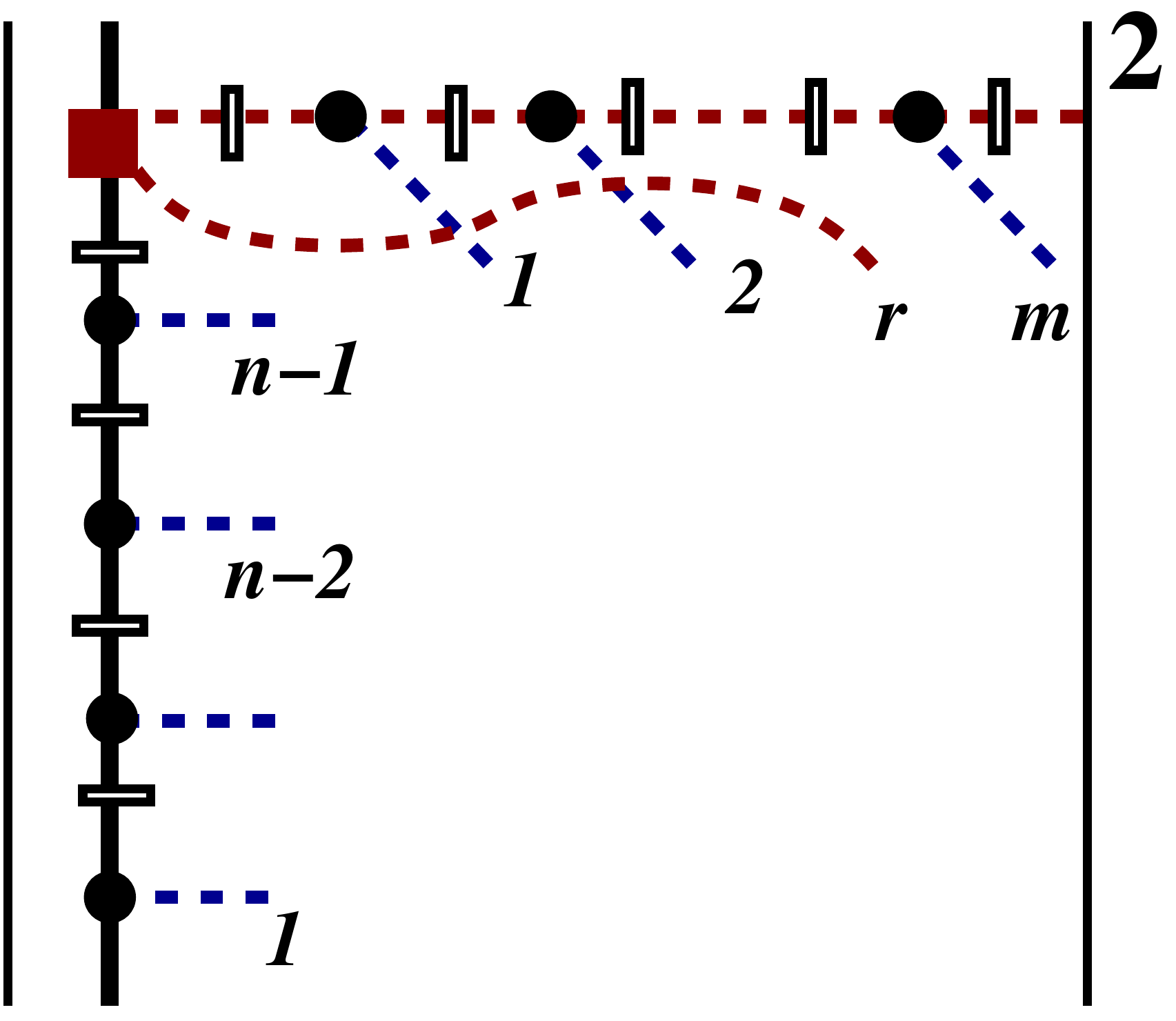}}
\Bigg\}
\\&
~
\\&
~\!\!\!\!\!\!\!\!\!\!\!\!
= e^{-S_{_{ISR}}}
\Bigg\{ \delta_{x=1}
+\sum_{n=1}^\infty\;
\bigg( \prod_{i=1}^n\; 
    \int\limits_{Q>a_i>a_{i-1}}\!\!\!\!\!\!  \frac{d^3 k_i}{k_i^0}\;
    \rho^{(1)}_{1B}(k_i)
\bigg)
e^{-S_{_{FSR}}}
\sum_{m=0}^\infty\;
\bigg( \prod_{j=1}^m\; 
    \int\limits_{Q>a_{nj}>a_{n(l-1)}}\!\!\!\!\!\! \frac{d^3 k'_j}{{k'}_j^{0}}
    \rho^{(1)}_{1V}(k'_j)
\bigg)
\\&~~~~~~~~~~~~~~~~~~~~~~~~~~~~~\times
\bigg[
\beta_0^{(1)}(z_n)
+\sum_{j=1}^{n-1} W(\tilde{k}_n, \tilde{k}_j)
+\sum_{r=1}^{m}   W(\tilde{k}_n, \tilde{k}'_r)
\bigg]
\delta_{x=\prod_{j=1}^n x_j}
\Bigg\},
\end{split}
\end{equation}
where the MC weights are given by
\begin{equation}
\beta_0^{(1)}\equiv \frac{%
 \left| \raisebox{-7pt}{\includegraphics[height=7mm]{xBrBet0ISR.pdf}}
 \right|^2}%
{ \left|\raisebox{-7pt}{\includegraphics[height=7mm]{xBrBorn.pdf}}
 \right|^2 },\quad
W(k_2,k_1)\equiv \frac{%
 \left| \raisebox{-7pt}{\includegraphics[height=7mm]{xBrBetISR.pdf}}
 \right|^2 }{%
 \left| \raisebox{-7pt}{\includegraphics[height=7mm]{xBr2ReCt.pdf}}
 \right|^2
+\left| \raisebox{-7pt}{\includegraphics[height=7mm]{xBr2gReCt.pdf}}
\right|^2
}= \frac{%
 \left| \raisebox{-7pt}{\includegraphics[height=7mm]{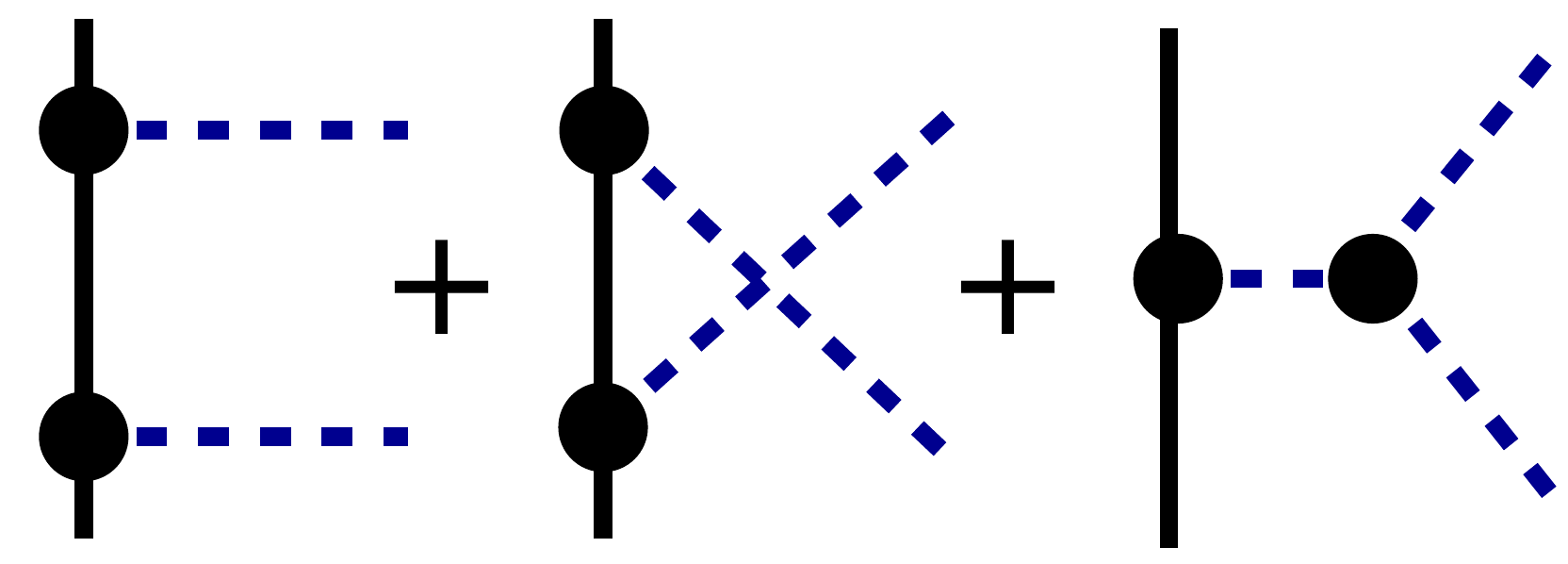}}
\right|^2 }{%
 \left| \raisebox{-7pt}{\includegraphics[height=7mm]{xBr2ReCt.pdf}}
 \right|^2
+\left| \raisebox{-7pt}{\includegraphics[height=7mm]{xBr2gReCt.pdf}}
\right|^2}\; -1,
\end{equation}
are finite, since all collinear and soft, ISR and FSR, singularities
are subtracted.
In particular, in the virtual correction
$
\left|
 \raisebox{-9pt}{\includegraphics[height=9mm]{xBrBet0ISR.pdf}}
\right|^2
=\big(1+2\Re(\Delta_{_{ISR}}+V_{_{FSR}} -S_{_{FSR}})\big)
\left|
  \raisebox{-9pt}{\includegraphics[height=9mm]{xBrBorn.pdf}}
\right|^2
$
FSR Sudakov $S_{_{FSR}}$ is subtracted.

Three-digit precision numerical test of the above FSR methodology
was done for a single NLO ISR+FSR insertion
for $n=1,2$ ISR gluons and infinite number of FSR gluons.
\begin{center}
\includegraphics[width=70mm,height=45mm]{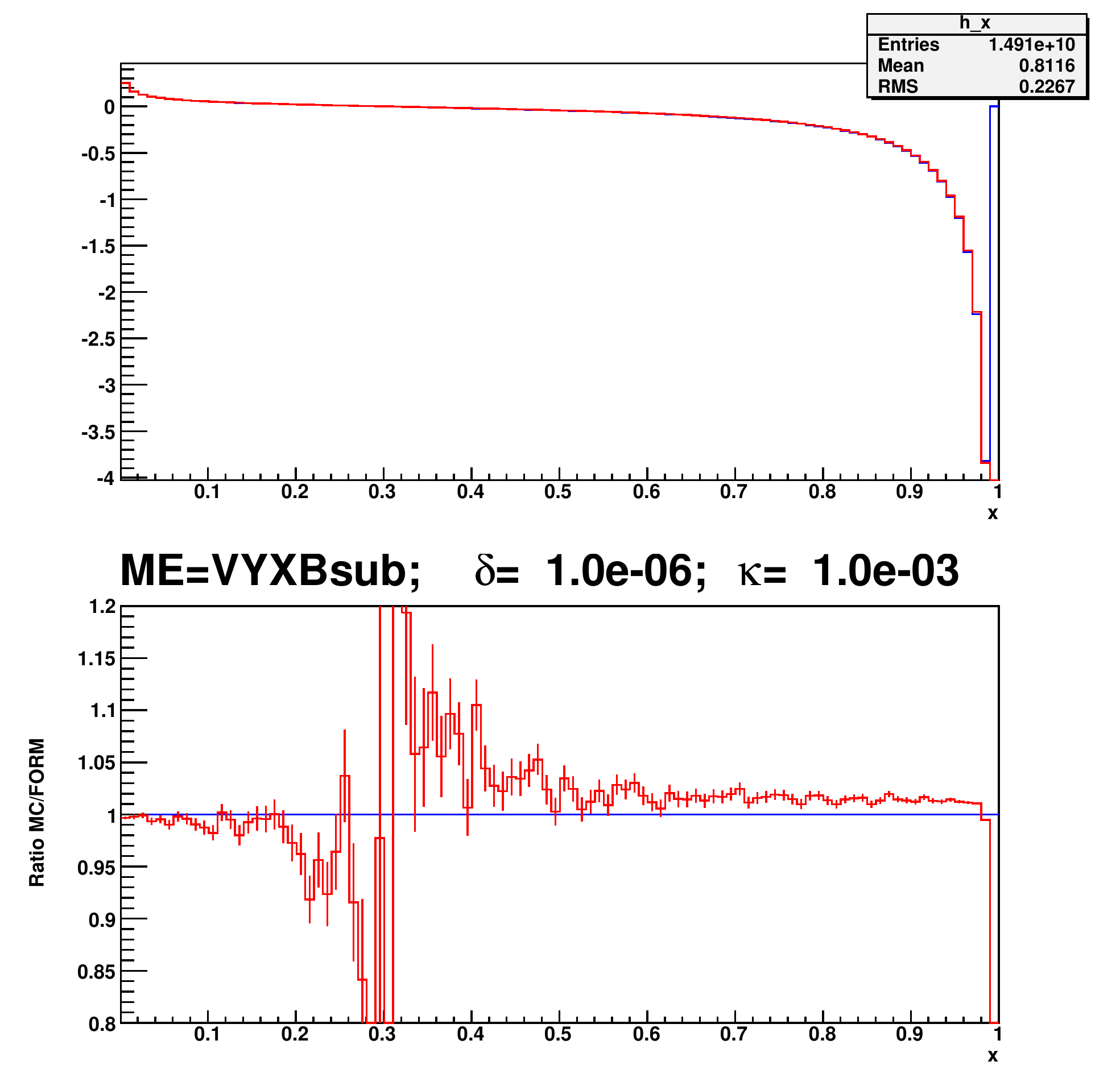}
\end{center}
The comparison is done between the MC and analytical results.

The exclusive approach presented here has been tested against an
independent MC implementing standard inclusive DGLAP evolution in
the NLO level. 
We can see the perfect agreement of both approaches
on the level of few promiles, see~\cite{Jadach:2010aa,Skrzypek:2009jk,Jadach:2009gm}.
Let us stress that in our exclusive MC scenario we have complete insight into the
transverse degrees of freedom up to NLO level, something that
is unavailable in other common inclusive or semi-inclusive approaches.
Our exclusive NLO MC features good behaviour of the MC weights --
weights are positive and narrowly distributed.
The techniques presented above are promising and will be exploited
in the construction of NLO MC generators for W/Z production @LHC and DIS @HERA.
Works related to this ongoing project are reported in refs.%
~\cite{Jadach:2010aa,Jadach:2009gm,Jadach:2010ew,Kusina:2010gp,Slawinska:2009gn}.

\end{document}